\documentclass[a4paper]{article}

\usepackage[pages=all, color=black, position={current page.south}, placement=bottom, scale=1, opacity=1, vshift=5mm]{background}

\usepackage[margin=.75in]{geometry} 

\usepackage{amsmath}
\usepackage{amsthm}
\usepackage{amssymb}

\usepackage[utf8]{inputenc}
\usepackage{hyperref}
\hypersetup{
	unicode,
	pdfauthor={Author One, Author Two, Author Three},
	pdftitle={A simple article template},
	pdfsubject={A simple article template},
	pdfkeywords={article, template, simple},
	pdfproducer={LaTeX},
	pdfcreator={pdflatex}
}

\usepackage[sort&compress,numbers,square]{natbib}
\theoremstyle{plain}

\theoremstyle{definition}

\usepackage{graphicx, color}
\graphicspath{{fig/}}

\usepackage{algorithm, algpseudocode} 
\usepackage{mathrsfs} 

\usepackage{lipsum}

\title{Athermal granular creep in a quenched sandpile}
\author{Nakul S. Deshpande$^1$ \and Paulo E. Arratia$^2$ \and Douglas J. Jerolmack$^2 , ^3$}

\date{
	$^1$Department of Physics, North Carolina State University \\ [2ex]
	$^2$Department of Mechanical Engineering and Applied Mechanics, University of Pennsylvania \\ [2ex]
 	$^3$Department of Earth and Environmental Science, University of Pennsylvania \\[2ex]
    corresponding author: \texttt{sediment@sas.upenn.edu}\\[2ex]%
}

\begin{document}
	\maketitle
	
	\begin{abstract}
 
		Creep is a generic descriptor of slow motions -- in the context of materials, it describes quasi-static deformation of a solid when subjected to stresses below the global yield, at which all rigidity collapses and the material flows. Here, we experimentally investigate creep, flow, and the transition between the two states in a granular heap flow. Within the surface flowing layer the dimensionless strain rate diminishes with depth, there is an absence of spatial correlations, and there is no aging dynamics. Beneath this layer, the bulk creeps via localized avalanches of plasticity, and there is significant aging. The transition between fast surface flow and slow bulk creep and aging is observed to be in the vicinity of a critical inertial number of $I = 10^{-5}$. Surprisingly, at the cessation of surface flow and the `quenching' of the pile, creep persists in the absence of the flowing layer; albeit with significant differences for a pile that experiences a long duration of surface flow (strongly annealed) and one where flow during preparation does not last long (weakly annealed). Our results contribute to an emerging view of athermal granular creep, showing similarities across dry and submerged systems. Quenched quiescent heaps that creep indefinitely, however, present a challenge to granular rheology, and open new possibilities for interpreting and casting creep and deformation of soils in nature. 
		
	\end{abstract}

	
\section{Introduction}







Slow motions of all types are known in the vernacular as `creep'. In the context of materials, creep is often a nuisance -- it can contribute to fatigue and failure. While common to many materials, creep in amorphous solids is perhaps best encapsulated in the context of glasses. Quenching a liquid produces a glass: an amorphous fragile material characterized by frozen-in structural disorder, creep, long relaxation times, heterogeneous dynamics, and susceptible to aging/rejuvenation effects \cite{falkMaterialsScienceFlow2007,zanottoGlassyStateMatter2017,scallietRejuvenationMemoryEffects2019}. A central question to understanding all of these dynamics is identifying the elementary unit responsible for plasticity. In amorphous solids, shear transformation zones \cite{falkDynamicsViscoplasticDeformation1998} are a candidate agent that plays a role akin to dislocations in crystalline solids. Both rely on the presence of temperature, which permits molecules or particles to overcome local energy barriers and thus manifests deformation of the bulk. Visualizing and measuring these dynamics in pure atomic glasses is a difficult experimental feat \cite{huangImagingAtomicRearrangements2013}. In lieu of purely thermal systems, workers have turned to other amorphous solids as analog materials -- including bubble rafts to model crystalline dislocations \cite{argonPlasticFlowDisordered1979} and colloidal ''glass'' suspensions \cite{jensenLocalShearTransformations2014}, where the thermal strength can be tuned and the system can be internally imaged. Shear transformation zones and creep persist even in quiescent, non-sheared colloidal suspensions by virtue of thermal motions \cite{jensenLocalShearTransformations2014}. Creep is observed in systems where global shear is minimized but the strength of temperature or other stresses increase \cite{berutBrownianGranularFlows2019}, and has recently been extended to fluid-driven athermal granular materials \cite{houssaisAthermalSedimentCreep2021}. We recently discovered that piles of athermal granular materials creep indefinitely, even in the absence of applied disturbance \cite{deshpandePerpetualFragilityCreeping2021}, exhibiting dynamical hallmarks of their thermal counterparts. However, further connections to existing views of granular rheology, heap flows, and glasses remain to be demonstrated. 

Athermal granular materials exhibit creep when driven; a well-studied source of driving is the avalanching of a hydrodynamic-like 'flowing layer' of grains, where collisions and dissipation dominate. When viewed over progressively longer exposure times, grain motions are not limited to this flowing layer; they appear to bleed into the bulk and creep amongst a set of long-lived neighbors \cite{komatsuCreepMotionGranular2001}. This transition between creeping and flow has been proposed to be a kind of jamming transition, where depth (and therefore confining stresses) replaces temperature as the control parameter \cite{katsuragiJammingGrowthDynamical2010}. Further insight into this transition came from a different physical system -- glass beads submerged in a viscous oil -- and identified a continuous transition between creep in the bulk and the flowing layer at the surface, albeit at a sharp boundary in the dimensionless shear rate \cite{houssaisOnsetSedimentTransport2015,houssaisRheologySedimentTransported2016}. This idea of treating creep and flow as a phase transition has intuitive appeal, occurs at a similar dimensionless shear rate, and displays some evidence of sharing similarities with second-order phase transitions \cite{ferdowsiGlassyDynamicsLandscape2018}. Creep can lead to compaction and the development of fabric anisotropy in grain arrangements, with \cite{cunezStrainHardeningSediment2022} or without \cite{biJammingShear2011} a flowing layer. This complicates the picture wherein the flowing layer plays the leading role of generating creep. Further, aging in the bulk has not been observed in dry heap flows, and while undisturbed piles exhibit sub-critical creep, there is no view that fully describes the departure of the system from fully-activated surface flow to a quenched and quiescent state where all driving is minimized.


Here, we seek to address a series of open questions regarding athermal granular creep and avalanching surface flow. Is there a difference between creep with a flowing layer and without? Does the flowing nature have consequences for creep in the bulk? What is the separation between creep and flow? How significantly do the dynamics change across the transition? To answer these questions we measure grain motions using spatially-resolved diffusing wave spectroscopy (DWS) \cite{amonSpatiallyResolvedMeasurements2017}, which allows us to simultaneously measure the dynamics of the flowing layer and those of the bulk.




\section{Experimental Setup and Methods}

\subsection{Experimental Setup}
Our experiments are conducted in a heap (0.0127 m W x 0.10 cm L) with two open outlets. The walls of the apparatus are made of borosilicate glass and the entire system sits on a vibration-isolated table. We conduct experiments of three kinds. The first is to supply grains (flux $Q = 2.5 g s^{-1}$)  into the cell for a duration of $10^{3}$ s, or about 16 minutes: this is the "flowing layer" experiment. There are then two `quenched' experiments: one which experiences a duration of $10^{3}$ s of supplied flux and another which experiences $1$ s of surface flow before cessation, these are the "strongly annealed" and "weakly annealed" experiments, respectively. Grains are soda-lime glass beads with $d = 100 \mu m$ and density $\rho = 2650 kg/m^{3}$.

\begin{figure}[ht]
\centering
\includegraphics[width=1\linewidth]{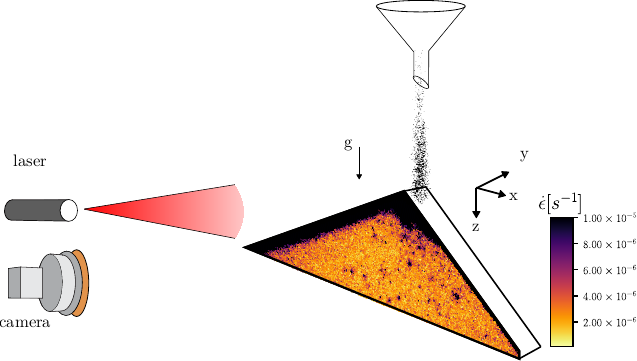}
\caption[Experimental setup]{\textbf{Experimental setup.}}
\label{fig:figure1}
\end{figure}



\subsection{Spatially-resolved Diffusive Wave Spectroscopy}

To measure grain motions, we implement spatially-resolved diffusive wave spectroscopy (DWS). \cite{deshpandePerpetualFragilityCreeping2021,amonSpatiallyResolvedMeasurements2017}. The advantage of spatially-resolved DWS is twofold: we are able to detect small-magnitude grain motions that are on the order of the optical wavelength, and we are able to generate significant statistics of strain and timescales within the flowing layer, the creeping bulk, and the quenched, undisturbed piles. It is a flexible and powerful technique. The principle of DWS is that highly coherent laser light is expanded and projected upon the heap. Photons scatter amongst the grains, interfere, and return a random interference pattern known as a 'speckle pattern'. As grains move, the speckle pattern changes. Grain motions can be accessed through an inverse process via analysis of the quantity of interest, $G$:

\begin{eqnarray*}
G(t_w ,t_w +\tau) = \frac{\left \langle i_{t_w} i_{t_w + \tau} \right \rangle - \left \langle i_{t_w} \right \rangle \left \langle i_{t_w + \tau} \right \rangle}{ \sqrt{\left \langle i_{t_w}^{2} \right \rangle-{\left \langle i_{t_w} \right \rangle}^{2}}{ \sqrt{\left \langle i_{t_w + \tau}^{2} \right \rangle-{\left \langle i_{t_w + \tau} \right \rangle}^{2}}}}
\end{eqnarray*}

\noindent where $i$ indicates the metapixel (see below) intensity, $\langle \rangle$ are spatial averages within a metapixel and \textit{c} is a constant which depends on the experimental parameters: $c = 8\pi \frac{\sqrt{2/5l^{*}}}{\lambda}$, where $\lambda$ is the wavelength of light. Thus, $G$ is a quantity which indicates the change in the speckle pattern within a metapixel between a start time $t$ and a lag time $\tau$. Using optical theory \cite{amonSpatiallyResolvedMeasurements2017}, we can map $G$ to a physical quantity: $\varepsilon$, the amount of strain which occurs inside a volume fixed by $l^{*}$: $G(t,\tau) = exp(-c(\overline{\varepsilon}+\overline{\zeta}))$. There are two contributions to changes in $G$: ${\varepsilon}$ is the strain and ${\zeta}$ are random motions.

For the flowing experiments, we set the frame rate to 100 Hz and compute 14 correlation functions, with waiting times, $t_{w}$, that are sampled at log-spaced intervals: (0,2,4,8,16...4016,8192 $10^{-2} s$), where the unit of time is set by the frame rate of the camera. The first waiting time $t_{w} = 0$ is set to be the instant during preparation that a fully-formed, space-filling wedge is formed and the volume of the heap does not change. Both annealed experiments are measured with a frame rate of 1 Hz, and the spacing of $t_w$ is (0,2,4,8,16...4016,8192 s). 

The quantity extracted from $G$ is an ensemble-averaged strain within a scattering volume whose size is set by the mean free path $l^{*}$ of the photons; this is about three grain diameters ($3.3 d$). In these experiments, a camera zooms and focuses in on about $4.5 cm$ of space. The pixel resolution of the camera is 2048 x 2048 pixels; yielding a resolution of about $2.2 x 10^{-5} m/px$ for the collected speckle images. The optics scaling are crucial, as they set the limits of our ability to achieve spatially-resolved measurements. From raw stacks of speckle images, we partition/coarse grain the field of view into `metapixels', where the grid size is selected to be $l^* = 3.3 d$, the mean-free path of photons and thus the scattering volume within which deformation occurs.




\subsection{Extracted quantities}

Our main observable in these experiments is the speckle correlation function $G$. As grains move, $G$ decorrelates from 1 to 0. We partition the region of interest into the flowing zone and the bulk beneath, and then compute the spatial average $\langle G \rangle$. A characteristic timescale that captures the dynamics responsible for the decorrelation can be extracted by fitting a stretched exponential function of the form: $G = exp \left [\frac{-t}{\tau_e} \right ]^\beta$ -- a common form for relaxation in glassy systems, amorphous solids, and granular heap flow \cite{yuStretchedExponentialRelaxation2015,lukichevPhysicalMeaningStretched2019,katsuragiJammingGrowthDynamical2010}. This fit is performed for all of the metapixels within an image. The quantity $\tau_e$ then gives a time for a given correlation function to achieve $1/e$ that corresponds to a strain using the relation above; this yields a strain rate $\dot{\epsilon}$. This allows us to compute the dimensionless strain rate, the inertial number:

\begin{equation}
  I  = \frac{\dot{\epsilon} d}{\sqrt{{\frac{P}{\rho}}} }
\end{equation}

\noindent where $d$ is the grain size, $P$ is the pressure and $\rho$ is the grain density. The inertial number is a comparison of the timescales of mean flow and of grain re-arrangements; it is in essence a shear P\'eclet-like number. To generate profiles of $I$, we fit the correlation within each metapixel, extract the timescale, compute the horizontally-averaged (x) value, and plot this value against depth $z$. We then convert to strain rate, and compute $I$ for each depth. In addition, we compute the spatial auto correlation of maps of the fitted timescale $\tau_{e}$. Spatial auto correlations are computed by selecting a rectangular region of interest of the strain-rate field, aligned at the free surface and computing the 2D Fourier Transform. The transform is then inverted.

\begin{figure}[ht]
\centering
\includegraphics[width=1\linewidth]{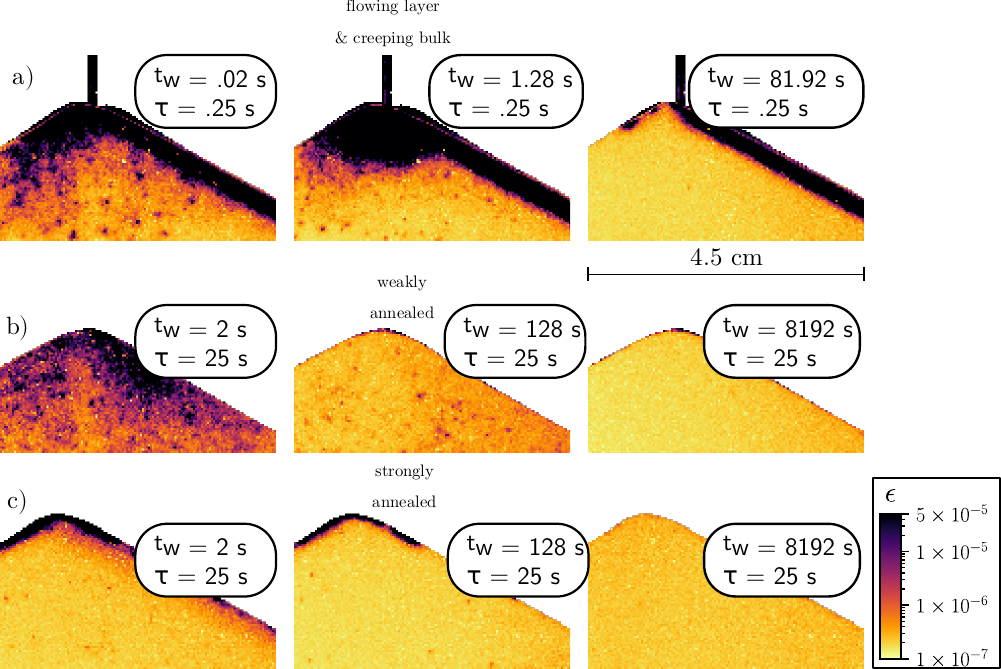}
\caption[Phenomenology: flowing, weakly annealed, strongly annealed]{\textbf{Phenomenology: flowing, weakly annealed, strongly annealed}. Heat maps of cumulative strain for the three types of experiments.}
\label{fig:ch4fig2}
\end{figure}

\section{Results}

\subsection{Phenomenology}

In a typical experimental run, avalanching grains begin by alternating between each of the two open outlets of the system, but eventually choose one of the two once the pile is fully formed. At early waiting times i.e. those close to the initial formation of the pile, the flowing layer appears as a zone of coherent flow riding above a bulk that features localized and intermittent zones of strain (Figure \ref{fig:ch4fig2}). As time progresses, this activity in the bulk diminishes, until it appears, for the same lag time, there is no bulk deformation at all. The addition of grains from above shows a radial zone beneath the apex of the pile, a likely affect of the free-fall of grains. 

At the cessation of flow, localized zones of deformation persist. In the case of a weakly annealed pile that only experiences $1 s$ of surface flow, this deformation is found throughout the pile, at the free surface and within the bulk, and gradually diminishes with waiting time. Strongly annealed piles that have been sheared by the flowing layer for $10^{3}$s display much less deformation within the bulk, and most motion is constrained to the free surface and at the apex of the pile. 

\begin{figure}[ht]
\centering
\includegraphics[width=1\linewidth]{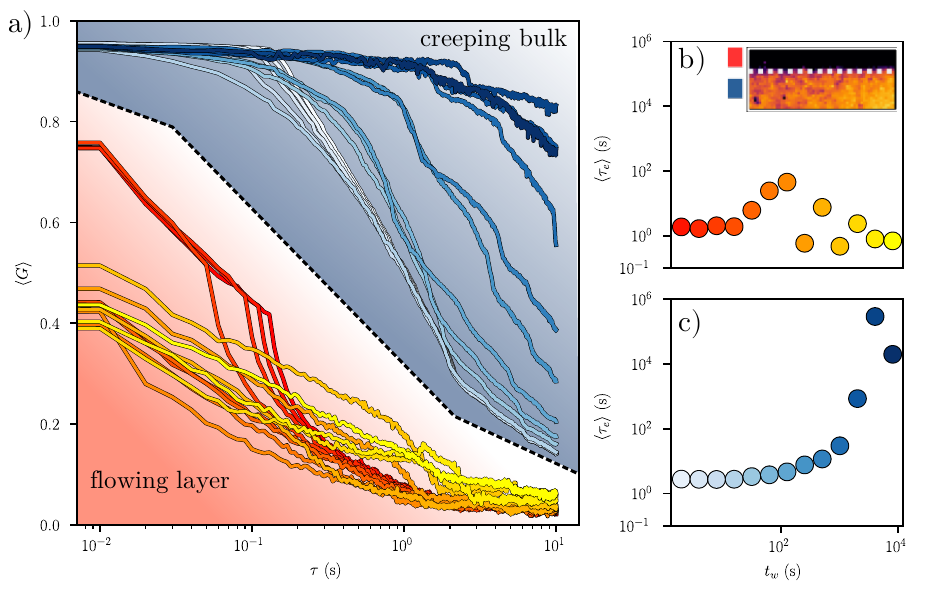}
\caption[Correlations in the flowing layer and in the creeping bulk]{\textbf{Correlations in the flowing layer (reds) and in the creeping bulk (blues).} Correlation functions are for 14 start times and spatially averaged across a region of interest that includes only metapixels within the flowing layer, or within the bulk. Characteristic relaxation times for each correlation function for the b) flowing layer c) and creeping bulk. The region of interest and separation of the flowing layer and the bulk is shown in b), separated by the dotted line. Note that the colors in b) and c) correspond to the waiting times shown in a). Note that in (b), $\langle \tau_{e} \rangle$ is more or less constant, while in (c) it increases rapidly. This indicates aging of creep, and no aging for the flow.}
 \label{fig:ch4fig3}
\end{figure}

\subsection{Correlation functions and timescales}

\begin{figure}[ht]
\centering
\includegraphics[width=1\linewidth]{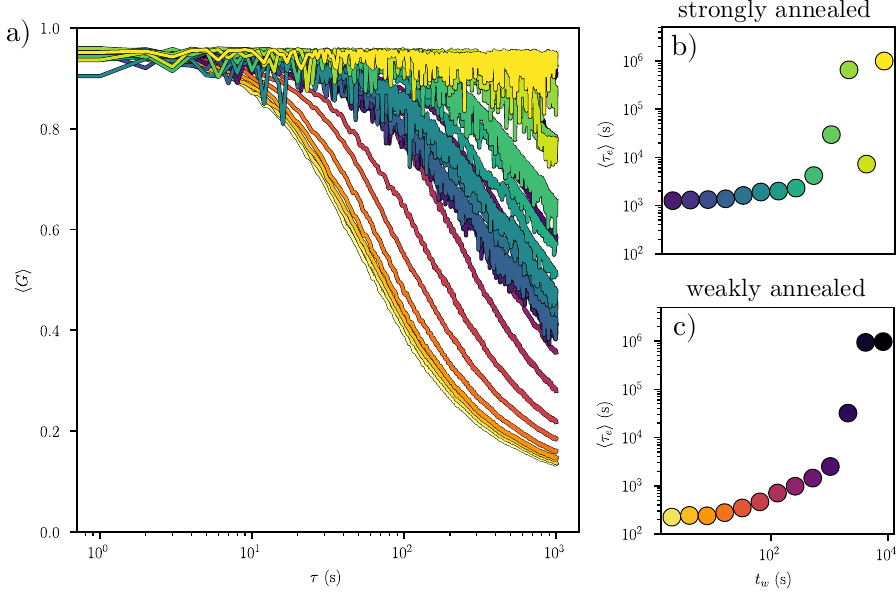}
\caption[Correlations in quenched and annealed piles]{\textbf{Correlations in quenched and annealed piles.} Correlation functions for 14 start times for a weakly annealed pile (inferno colormap) and strongly annealed pile (viridis colormap). Characteristic relaxation times for each type of pile}
\label{fig:ch4fig4}
\end{figure}

In the flowing layer, correlations decay quickly (Figure \ref{fig:ch4fig3}a) These dynamics are relatively constant in time; the characteristic timescale for each waiting time $t_{w}$ is bounded between $10^{0}$ and $10^{2}$s. Correlations in the creeping bulk grow systematically with $t_{w}$ from $10^{0}$s to $10^{6}$s (Figure \ref{fig:ch4fig3}c). This supports the qualitative picture illustrated above, where deformation in the bulk diminishes with time. At the cessation of flow, quenched piles continue to decorrelate when strongly and weakly annealed (\ref{fig:ch4fig4}a). In the weakly annealed case, this decorrelation begins quickly, slows with time (Figure 3 \ref{fig:ch4fig4}b), and exhibits smooth relaxation. In the strongly annealed case, the initial relaxation times are longer and are plateaued around $10^{3}$s before increasing (Figure 3 \ref{fig:ch4fig4}c). These correlations also feature a large degree of noise superimposed onto the gradual relaxation. 

\subsection{Profiles and spatial correlations}

\begin{figure}[ht]
\centering
\includegraphics[width=1\linewidth]{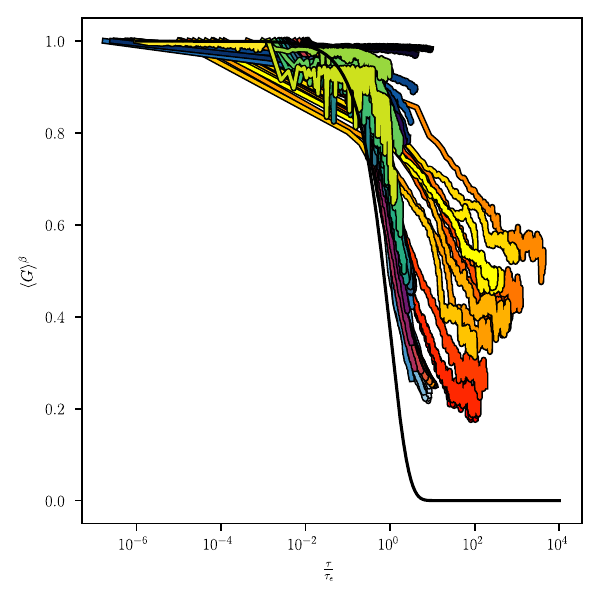}
\caption[Collapse of all correlation functions]{\textbf{Collapse of all correlation functions.} All correlation functions reasonably collapsed by parameters extracted from stretched exponential fit. The collapse is poor for the correlation functions of the flowing layer; this may be due to camera frame rates that are too coarse to accurately determine the fit parameters.}
\label{fig:ch4fig5}
\end{figure}

Horizontally-averaged profiles of $I$ for the flowing layer rapidly decrease with depth (Figure \ref{fig:ch4fig5}a). Flowing layer profiles do not change appreciably with waiting time. In the range of $I = 10^{-5} - 10^{-7}$, there is a sharp kink in the profiles. This kink demarcates fast flow with the slowly creeping bulk. Unlike the flowing layer, profiles of $I$ show a strong time-dependent behavior, and with increasing waiting time the horizontally-averaged value of $I$ decreases from $10^{-8}$ to $10^{-13}$; a seven order of magnitude change (Figure \ref{fig:ch4fig5}a). Both weakly and strongly annealed piles exhibit time-dependent behavior, as $I$ decreases with $t_{w}$ (Figure \ref{fig:ch4fig5}b and c).

\begin{figure}[ht]
\centering
\includegraphics[width=1\linewidth]{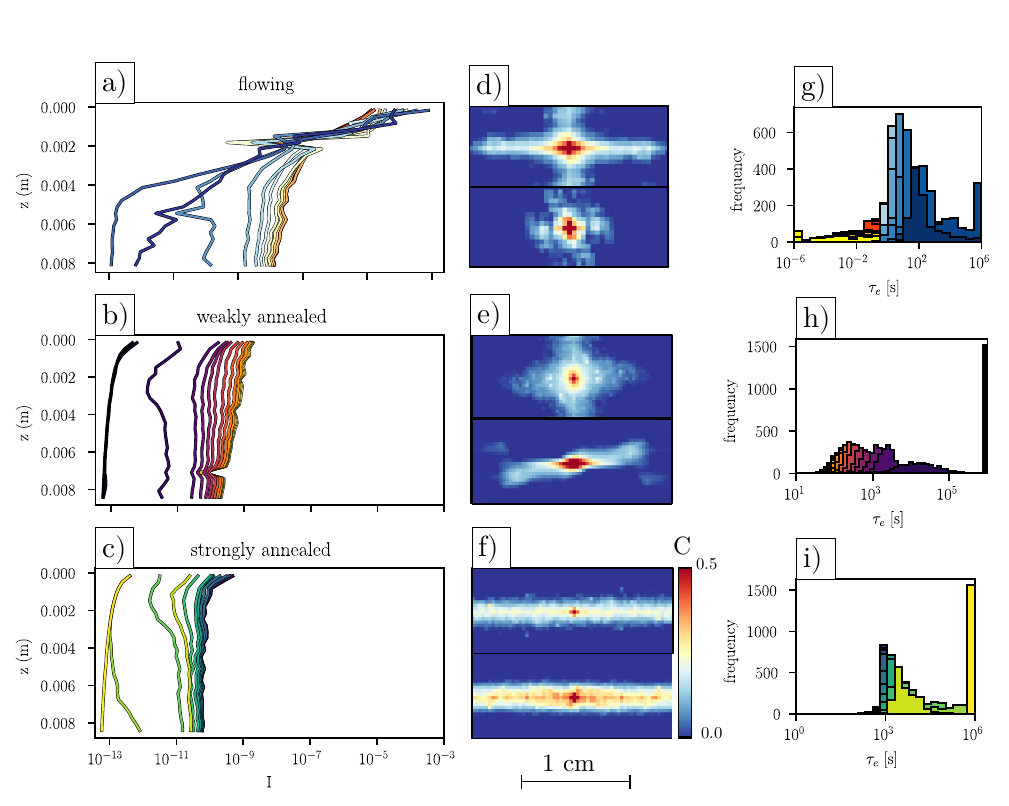}
\caption[Horizontally-averaged profiles, spatial correlations, and relaxation timescale distributions]{\textbf{Horizontally-averaged profiles, spatial correlations, and relaxation timescale distributions.} In the flowing layer experiments, the depth profiles show no aging in the top flowing layer and aging in the creep layer, separated by $I = 10^-5$ a). Weakly and strongly annealed depth profiles (b) and c)) age significantly -- note that the weakly annealed profiles begin at larger magnitude $I$ values than the strongly annealed case. Spatial correlations of the strain field within the creeping bulk show four-fold symmetric patterns -- known as `quadrupoles' (d,e,f). The degree of symmetry changes during the course of an experiment, and each panel corresponds to a strain field at an arbitrary time to illustrate the changes in the features of the correlation. It is notable that for the strongly annealed experiments f), the correlations are highly anisotropic, which may be an expression of grain fabrics developed during the long period (about 16 minutes) of prepatory surface flow. Histograms of the intrinsic relaxation time $\tau_e$ show distinct populations for each experiment (g,h,i). These stay clustered in the flowing layer and indicate little aging g), while for quenched experiments h), i), the distributions are stretched to lower timescales, indicating aging. 
}
 \label{fig:ch4fig6}
\end{figure}

Spatial correlations of the extracted timescales display distinct and coherent structure (Figure \ref{fig:ch4fig6}d). The creeping bulk beneath the flowing layer and the weakly annealed pile show four-fold symmetric correlations, characteristic of Eshelby inclusions in deformed amorphous solids \cite{nicolasDeformationFlowAmorphous2018}. This symmetry is broken at later $t_{w}$ for the weakly annealed pile, and is strongly anisotropic in the correlations of the strongly annealed pile. 

Histograms of the fitted relaxation timescales emphasizes the time-independence of the flowing layer, and the time dependence of creep beneath flow, and in the annealed, quenched experiments (\ref{fig:ch4fig5} a,b,c). Flowing layer timescales are fast and cluster near the lower limit of $10^{-6}$s. Timescale distributions in the bulk stretch and smear, indicating significant aging. Weakly annealed distributions likewise show significant aging, while the strongly annealed pile does not. 

\section{Discussion}

We have shown that deformation in a granular heap flow is not restricted to the hydrodynamic flowing layer, there is pervasive creep within the bulk as well. This creep is not steady, its magnitude decreases in time and therefore ages. Despite this aging, the boundary between creep in the bulk and flow at the surface occurs in the vicinity of $I = 10^{-5}$. This is significant, because the same value of $I$ has also been shown to separate creep and flow in a different experimental setting of fluid-driven sediment transport \cite{houssaisOnsetSedimentTransport2015} and in dry DEM simulations \cite{ferdowsiGlassyDynamicsLandscape2018}. Perhaps even more surprisingly, this flowing layer is not required for creep to exist. Quenched piles, where the flow is completely removed, also exhibit time-varying creep and aging. An important point is that this aging is strongly connected to the duration of applied surface flow prior to cessation -- the imprint of aging during flow has consequences for the initial magnitude of relaxation as well as its functional form. 

Spatial correlations of the intrinsic relaxation timescales exhibit strong structure. Four-fold symmetric quadrupolar spatial correlations are a mesoscopic signature of plastic deformation in amorphous solids and glasses \cite{nicolasDeformationFlowAmorphous2018}. These signatures are typically found in thermal systems, and most notably in a colloidal suspension in the absence of shear \cite{jensenLocalShearTransformations2014}. It is therefore interesting that these correlations exhibit Eshelby-like quadrupolar shapes, given that our system is athermal. Anisotropic spatial correlations may hint at an underlying structure within the grains -- and a reflection of force-chain fabric and preferential alignment as observed in shear-jammed granular materials \cite{biJammingShear2011}. A paradoxical observation is the \textit{increase} of the inertial number with depth once the flow is removed and the system is quenched. This suggests grains slipping between the interface of the glass walls and the bottom boundary and is visible in strain maps at late $t_{w}$. Indeed, we have shown elsewhere the extreme sensitivity of quenched creep glassy dynamics in systems where the bottom boundary is bound. Note that DWS is really imaging a few grain diameters from the wall; the heightened strain rates could be a reflection of a boundary effect of the apparatus.

There are some issues with the present experiments that should be incorporated into future experiments to improve and clarify the above results. First, the system should be restricted to only one outlet to ensure that all of the supplied flux is transported in the region of interest over which the horizontally-averaged profiles are computed. Otherwise, unsteady flow could give the impression of aging dynamics that are not actually present. Second, we note that there is significant variability that qualitatively depends on the relative humidity in the laboratory -- the correlations are qualitatively rougher and slower with RH  \% 50 (experiments presented here) than in conditions with RH \% 30. Future work should seek to make these changes to the experimental apparatus, and make a more formal comparison between humidity conditions. 

\clearpage

\paragraph{Acknowledgements} 

We acknowledge the funding provided by the Army Research Office (ARO; Grant W911NF2010113) and NSF Materials Research Science and Engineering Center (NSF-DMR-1720530) to D.J.J. and P.A., and the Petroleum Research Fund (ACS-PRF Grant 61536-ND8) to D.J.J.

\bibliography{main.bib} 


\end{document}